\documentstyle[12pt,epsfig]{article}
\def\t0{t$_{0}$}
\def\ecm{$\sqrt{s}$}
\def\ptac{$0.0375 \sqrt{s}$}
\def\cs{$\cos{\theta}$}

\def\cost{${\mathrm  \left|\cos\theta\right| < 0.95}$}
\def\cost2{${\mathrm \left|\cos\theta\right| < 0.8}$}
\def\cosa{${\mathrm \cos{\alpha} > 0.98}$}
\def\gev{\mbox{GeV}}
\def\gevsq{\mbox{GeV/$c^{2}$}}

\def\bha{${\mathrm e^{+}e^{-}\rightarrow e^{+}e^{-}}$}
\def\nunug{${\mathrm \nu \bar{\nu} \gamma (\gamma)}$}
\def\nunugg{${\mathrm \nu \bar{\nu} \gamma \gamma}$}
\def\nunuggg{${\mathrm \nu \bar{\nu} \gamma \gamma (\gamma)}$}
\def\ggg{${\mathrm \gamma \gamma (\gamma)}$}
\def\cdfg{${\mathrm q \bar{q} \rightarrow \tilde{e} \tilde{e} \rightarrow ee \chi_{1}^{0} \chi_{1}^{0} \rightarrow ee \tilde{G} \tilde{G} \gamma \gamma}$}
\def\cdfx{${\mathrm q \bar{q} \rightarrow \tilde{e} \tilde{e} \rightarrow ee \chi_{2}^{0} \chi_{2}^{0} \rightarrow ee \chi_{1}^{0} \chi_{1}^{0}  \gamma \gamma}$}
\def\lepx{${\mathrm e^{+}e^{-} \rightarrow \chi_{2}^{0} \chi_{2}^{0} \rightarrow \chi_{1}^{0} \chi_{1}^{0} \gamma \gamma}$}
\def\lx{${\mathrm e^{+}e^{-} \rightarrow \chi_{2}^{0} \chi_{1}^{0} \rightarrow \chi_{1}^{0} \chi_{1}^{0} \gamma }$}
\def\lepg{${\mathrm e^{+}e^{-} \rightarrow \chi_{1}^{0} \chi_{1}^{0} \rightarrow \tilde{G} \tilde{G} \gamma \gamma}$}
\def\lepgx{${\mathrm e^{+}e^{-} \rightarrow \chi_{1}^{0} \tilde{G} \rightarrow \tilde{G} \tilde{G} \gamma }$}
\def\lep1{${\mathrm e^{+}e^{-} \rightarrow \chi_{1}^{0} \chi_{1}^{0} }$}
\def\eenunu{${\mathrm e^{+}e^{-}\rightarrow \nu \bar{\nu} \gamma (\gamma)}$}

\def\eegg{${\mathrm e^{+}e^{-}\rightarrow   \gamma \gamma (\gamma)}$}
\def\xtwo{${\mathrm \chi_{2}^{0}}$}
\def\xone{${\mathrm \chi_{1}^{0}}$}

\def\xxg{${\mathrm \chi_{2}^{0} \rightarrow \chi_{1}^{0} \gamma }$}
\def\eex2x2{${\mathrm e^{+}e^{-}\rightarrow \chi_{2}^{0} \chi_{2}^{0} \rightarrow 
\chi_{1}^{0}  \chi_{1}^{0} \gamma \gamma }$}
\def\pb{${\mathrm pb^{-1}}$}

\begin{document}
\begin{center}
\centerline{\Large EUROPEAN LABORATORY FOR PARTICLE PHYSICS (CERN)}
\end{center}

\vspace{0.5cm}

\begin{flushright}
CERN-PPE/97-122 \\
September 4 1997 \\
\end{flushright}

\thispagestyle{empty}

\vspace{1.5cm}

\begin{center}
{\LARGE \bf Searches for Supersymmetry in the photon(s) plus missing energy channels at $\sqrt{s}$ = 161 GeV and 172 GeV}

\end{center}

\vspace{1.0cm}

\begin{center}
{\large The ALEPH Collaboration$^{*}$}
\end{center}

\vspace{2.5cm}

\begin{abstract}
 Searches for supersymmetric particles in  channels with one or more photons and missing energy have been performed with data collected by
the ALEPH detector at LEP. The data consist of 11.1 \pb\ at $\sqrt{s} = 161 ~\, \rm GeV$, 1.1 \pb\ at 170 \gev\ and 9.5 \pb\ at 172 GeV.   The \eenunu\ cross section is measured.  The data are in good agreement with predictions based on the Standard Model,  and are 
used to set upper limits on the cross sections for anomalous photon production. These limits are compared to two different SUSY models  and  used to set limits on the neutralino mass.
 A limit of 71 \gevsq\ at 95\% C.L. is set on the mass of the lightest neutralino ($\tau_{\chi_{1}^{0}} \leq $ 3 ns) for the
gauge-mediated supersymmetry breaking and LNZ models.
\end{abstract}

\vspace{2.5cm}

\begin{center}
{\sl (To be submitted to Physics Letters B)}
\end{center}

\mbox{}

\vspace{0.5cm}

\noindent
{\small {$^{*}\,${See the following pages for the list of authors.}} }\\

\pagebreak

\pagestyle{empty}
\newpage
\small
%
%
\newlength{\saveparskip}
\newlength{\savetextheight}
\newlength{\savetopmargin}
\newlength{\savetextwidth}
\newlength{\saveoddsidemargin}
\newlength{\savetopsep}
\setlength{\saveparskip}{\parskip}
\setlength{\savetextheight}{\textheight}
\setlength{\savetopmargin}{\topmargin}
\setlength{\savetextwidth}{\textwidth}
\setlength{\saveoddsidemargin}{\oddsidemargin}
\setlength{\savetopsep}{\topsep}
%
%
\setlength{\parskip}{0.0cm}
\setlength{\textheight}{25.0cm}
\setlength{\topmargin}{-1.5cm}
\setlength{\textwidth}{16 cm}
\setlength{\oddsidemargin}{-0.0cm}
\setlength{\topsep}{1mm}
\pretolerance=10000
\centerline{\large\bf The ALEPH Collaboration}
\footnotesize
\vspace{0.5cm}
{\raggedbottom
\begin{sloppypar}
\samepage\noindent
R.~Barate,
D.~Buskulic,
D.~Decamp,
P.~Ghez,
C.~Goy,
J.-P.~Lees,
A.~Lucotte,
M.-N.~Minard,
J.-Y.~Nief,
B.~Pietrzyk
\nopagebreak
\begin{center}
\parbox{15.5cm}{\sl\samepage
Laboratoire de Physique des Particules (LAPP), IN$^{2}$P$^{3}$-CNRS,
74019 Annecy-le-Vieux Cedex, France}
\end{center}\end{sloppypar}
\vspace{2mm}
\begin{sloppypar}
\noindent
M.P.~Casado,
M.~Chmeissani,
P.~Comas,
J.M.~Crespo,
M.~Delfino, 
E.~Fernandez,
M.~Fernandez-Bosman,
Ll.~Garrido,$^{15}$
A.~Juste,
M.~Martinez,
G.~Merino,
R.~Miquel,
Ll.M.~Mir,
C.~Padilla,
I.C.~Park,
A.~Pascual,
J.A.~Perlas,
I.~Riu,
F.~Sanchez
\nopagebreak
\begin{center}
\parbox{15.5cm}{\sl\samepage
Institut de F\'{i}sica d'Altes Energies, Universitat Aut\`{o}noma
de Barcelona, 08193 Bellaterra (Barcelona), Spain$^{7}$}
\end{center}\end{sloppypar}
\vspace{2mm}
\begin{sloppypar}
\noindent
A.~Colaleo,
D.~Creanza,
M.~de~Palma,
G.~Gelao,
G.~Iaselli,
G.~Maggi,
M.~Maggi,
N.~Marinelli,
S.~Nuzzo,
A.~Ranieri,
G.~Raso,
F.~Ruggieri,
G.~Selvaggi,
L.~Silvestris,
P.~Tempesta,
A.~Tricomi,$^{3}$
G.~Zito
\nopagebreak
\begin{center}
\parbox{15.5cm}{\sl\samepage
Dipartimento di Fisica, INFN Sezione di Bari, 70126
Bari, Italy}
\end{center}\end{sloppypar}
\vspace{2mm}
\begin{sloppypar}
\noindent
X.~Huang,
J.~Lin,
Q. Ouyang,
T.~Wang,
Y.~Xie,
R.~Xu,
S.~Xue,
J.~Zhang,
L.~Zhang,
W.~Zhao
\nopagebreak
\begin{center}
\parbox{15.5cm}{\sl\samepage
Institute of High-Energy Physics, Academia Sinica, Beijing, The People's
Republic of China$^{8}$}
\end{center}\end{sloppypar}
\vspace{2mm}
\begin{sloppypar}
\noindent
D.~Abbaneo,
R.~Alemany,
A.O.~Bazarko,$^{1}$
U.~Becker,
P.~Bright-Thomas,
M.~Cattaneo,
F.~Cerutti,
G.~Dissertori,
H.~Drevermann,
R.W.~Forty,
M.~Frank,
F.~Gianotti,
R.~Hagelberg,
J.B.~Hansen,
J.~Harvey,
P.~Janot,
B.~Jost,
E.~Kneringer,
I.~Lehraus,
P.~Mato,
A.~Minten,
L.~Moneta,
A.~Pacheco,
J.-F.~Pusztaszeri,$^{20}$
F.~Ranjard,
G.~Rizzo,
L.~Rolandi,
D.~Rousseau,
D.~Schlatter,
M.~Schmitt,
O.~Schneider,
W.~Tejessy,
F.~Teubert,
I.R.~Tomalin,
H.~Wachsmuth,
A.~Wagner$^{21}$
\nopagebreak
\begin{center}
\parbox{15.5cm}{\sl\samepage
European Laboratory for Particle Physics (CERN), 1211 Geneva 23,
Switzerland}
\end{center}\end{sloppypar}
\vspace{2mm}
\begin{sloppypar}
\noindent
Z.~Ajaltouni,
A.~Barr\`{e}s,
C.~Boyer,
A.~Falvard,
C.~Ferdi,
P.~Gay,
C~.~Guicheney,
P.~Henrard,
J.~Jousset,
B.~Michel,
S.~Monteil,
J-C.~Montret,
D.~Pallin,
P.~Perret,
F.~Podlyski,
J.~Proriol,
P.~Rosnet,
J.-M.~Rossignol
\nopagebreak
\begin{center}
\parbox{15.5cm}{\sl\samepage
Laboratoire de Physique Corpusculaire, Universit\'e Blaise Pascal,
IN$^{2}$P$^{3}$-CNRS, Clermont-Ferrand, 63177 Aubi\`{e}re, France}
\end{center}\end{sloppypar}
\vspace{2mm}
\begin{sloppypar}
\noindent
T.~Fearnley,
J.D.~Hansen,
J.R.~Hansen,
P.H.~Hansen,
B.S.~Nilsson,
B.~Rensch,
A.~W\"a\"an\"anen
\begin{center}
\parbox{15.5cm}{\sl\samepage
Niels Bohr Institute, 2100 Copenhagen, Denmark$^{9}$}
\end{center}\end{sloppypar}
\vspace{2mm}
\begin{sloppypar}
\noindent
G.~Daskalakis,
A.~Kyriakis,
C.~Markou,
E.~Simopoulou,
A.~Vayaki
\nopagebreak
\begin{center}
\parbox{15.5cm}{\sl\samepage
Nuclear Research Center Demokritos (NRCD), Athens, Greece}
\end{center}\end{sloppypar}
\vspace{2mm}
\begin{sloppypar}
\noindent
A.~Blondel,
J.C.~Brient,
F.~Machefert,
A.~Roug\'{e},
M.~Rumpf,
A.~Valassi,$^{6}$
H.~Videau
\nopagebreak
\begin{center}
\parbox{15.5cm}{\sl\samepage
Laboratoire de Physique Nucl\'eaire et des Hautes Energies, Ecole
Polytechnique, IN$^{2}$P$^{3}$-CNRS, 91128 Palaiseau Cedex, France}
\end{center}\end{sloppypar}
\vspace{2mm}
\begin{sloppypar}
\noindent
T.~Boccali,
E.~Focardi,
G.~Parrini,
K.~Zachariadou
\nopagebreak
\begin{center}
\parbox{15.5cm}{\sl\samepage
Dipartimento di Fisica, Universit\`a di Firenze, INFN Sezione di Firenze,
50125 Firenze, Italy}
\end{center}\end{sloppypar}
\vspace{2mm}
\begin{sloppypar}
\noindent
R.~Cavanaugh,
M.~Corden,
C.~Georgiopoulos,
T.~Huehn,
D.E.~Jaffe
\nopagebreak
\begin{center}
\parbox{15.5cm}{\sl\samepage
Supercomputer Computations Research Institute,
Florida State University,
Tallahassee, FL 32306-4052, USA $^{13,14}$}
\end{center}\end{sloppypar}
\vspace{2mm}
\begin{sloppypar}
\noindent
A.~Antonelli,
G.~Bencivenni,
G.~Bologna,$^{4}$
F.~Bossi,
P.~Campana,
G.~Capon,
D.~Casper,
V.~Chiarella,
G.~Felici,
P.~Laurelli,
G.~Mannocchi,$^{5}$
F.~Murtas,
G.P.~Murtas,
L.~Passalacqua,
M.~Pepe-Altarelli
\nopagebreak
\begin{center}
\parbox{15.5cm}{\sl\samepage
Laboratori Nazionali dell'INFN (LNF-INFN), 00044 Frascati, Italy}
\end{center}\end{sloppypar}
\vspace{2mm}
\begin{sloppypar}
\noindent
L.~Curtis,
S.J.~Dorris,
A.W.~Halley,
I.G.~Knowles,
J.G.~Lynch,
V.~O'Shea,
C.~Raine,
J.M.~Scarr,
K.~Smith,
P.~Teixeira-Dias,
A.S.~Thompson,
E.~Thomson,
F.~Thomson,
R.M.~Turnbull
\nopagebreak
\begin{center}
\parbox{15.5cm}{\sl\samepage
Department of Physics and Astronomy, University of Glasgow, Glasgow G12
8QQ,United Kingdom$^{10}$}
\end{center}\end{sloppypar}
\vspace{2mm}
\begin{sloppypar}
\noindent
O.~Buchm\"uller,
S.~Dhamotharan,
C.~Geweniger,
G.~Graefe,
P.~Hanke,
G.~Hansper,
V.~Hepp,
E.E.~Kluge,
A.~Putzer,
J.~Sommer,
K.~Tittel,
S.~Werner,
M.~Wunsch
\begin{center}
\parbox{15.5cm}{\sl\samepage
Institut f\"ur Hochenergiephysik, Universit\"at Heidelberg, 69120
Heidelberg, Fed.\ Rep.\ of Germany$^{16}$}
\end{center}\end{sloppypar}
\vspace{2mm}
\begin{sloppypar}
\noindent
R.~Beuselinck,
D.M.~Binnie,
W.~Cameron,
P.J.~Dornan,
M.~Girone,
S.~Goodsir,
E.B.~Martin,
P.~Morawitz,
A.~Moutoussi,
J.~Nash,
J.K.~Sedgbeer,
P.~Spagnolo,
A.M.~Stacey,
M.D.~Williams
\nopagebreak
\begin{center}
\parbox{15.5cm}{\sl\samepage
Department of Physics, Imperial College, London SW7 2BZ,
United Kingdom$^{10}$}
\end{center}\end{sloppypar}
\vspace{2mm}
\begin{sloppypar}
\noindent
V.M.~Ghete,
P.~Girtler,
D.~Kuhn,
G.~Rudolph
\nopagebreak
\begin{center}
\parbox{15.5cm}{\sl\samepage
Institut f\"ur Experimentalphysik, Universit\"at Innsbruck, 6020
Innsbruck, Austria$^{18}$}
\end{center}\end{sloppypar}
\vspace{2mm}
\begin{sloppypar}
\noindent
A.P.~Betteridge,
C.K.~Bowdery,
P.G.~Buck,
P.~Colrain,
G.~Crawford,
A.J.~Finch,
F.~Foster,
G.~Hughes,
R.W.L.~Jones,
T.~Sloan,
E.P.~Whelan,
M.I.~Williams
\nopagebreak
\begin{center}
\parbox{15.5cm}{\sl\samepage
Department of Physics, University of Lancaster, Lancaster LA1 4YB,
United Kingdom$^{10}$}
\end{center}\end{sloppypar}
\vspace{2mm}
\begin{sloppypar}
\noindent
I.~Giehl,
C.~Hoffmann,
K.~Jakobs,
K.~Kleinknecht,
G.~Quast,
B.~Renk,
E.~Rohne,
H.-G.~Sander,
P.~van~Gemmeren,
C.~Zeitnitz
\nopagebreak
\begin{center}
\parbox{15.5cm}{\sl\samepage
Institut f\"ur Physik, Universit\"at Mainz, 55099 Mainz, Fed.\ Rep.\
of Germany$^{16}$}
\end{center}\end{sloppypar}
\vspace{2mm}
\begin{sloppypar}
\noindent
J.J.~Aubert,
C.~Benchouk,
A.~Bonissent,
G.~Bujosa,
J.~Carr,
P.~Coyle,
C.~Diaconu,
A.~Ealet,
D.~Fouchez,
N.~Konstantinidis,
O.~Leroy,
F.~Motsch,
P.~Payre,
M.~Talby,
A.~Sadouki,
M.~Thulasidas,
A.~Tilquin,
K.~Trabelsi
\nopagebreak
\begin{center}
\parbox{15.5cm}{\sl\samepage
Centre de Physique des Particules, Facult\'e des Sciences de Luminy,
IN$^{2}$P$^{3}$-CNRS, 13288 Marseille, France}
\end{center}\end{sloppypar}
\vspace{2mm}
\begin{sloppypar}
\noindent
M.~Aleppo, 
M.~Antonelli,
F.~Ragusa
\nopagebreak
\begin{center}
\parbox{15.5cm}{\sl\samepage
Dipartimento di Fisica, Universit\`a di Milano e INFN Sezione di
Milano, 20133 Milano, Italy.}
\end{center}\end{sloppypar}
\vspace{2mm}
\begin{sloppypar}
\noindent
R.~Berlich,
W.~Blum,
V.~B\"uscher,
H.~Dietl,
G.~Ganis,
C.~Gotzhein,
H.~Kroha,
G.~L\"utjens,
G.~Lutz,
W.~M\"anner,
H.-G.~Moser,
R.~Richter,
A.~Rosado-Schlosser,
S.~Schael,
R.~Settles,
H.~Seywerd,
R.~St.~Denis,
H.~Stenzel,
W.~Wiedenmann,
G.~Wolf
\nopagebreak
\begin{center}
\parbox{15.5cm}{\sl\samepage
Max-Planck-Institut f\"ur Physik, Werner-Heisenberg-Institut,
80805 M\"unchen, Fed.\ Rep.\ of Germany\footnotemark[16]}
\end{center}\end{sloppypar}
\vspace{2mm}
\begin{sloppypar}
\noindent
J.~Boucrot,
O.~Callot,$^{12}$
S.~Chen,
A.~Cordier,
M.~Davier,
L.~Duflot,
J.-F.~Grivaz,
Ph.~Heusse,
A.~H\"ocker,
A.~Jacholkowska,
M.~Jacquet,
D.W.~Kim,$^{2}$
F.~Le~Diberder,
J.~Lefran\c{c}ois,
A.-M.~Lutz,
I.~Nikolic,
M.-H.~Schune,
L.~Serin,
S.~Simion,
E.~Tournefier,
J.-J.~Veillet,
I.~Videau,
D.~Zerwas
\nopagebreak
\begin{center}
\parbox{15.5cm}{\sl\samepage
Laboratoire de l'Acc\'el\'erateur Lin\'eaire, Universit\'e de Paris-Sud,
IN$^{2}$P$^{3}$-CNRS, 91405 Orsay Cedex, France}
\end{center}\end{sloppypar}
\vspace{2mm}
\begin{sloppypar}
\noindent
\samepage
P.~Azzurri,
G.~Bagliesi,$^{12}$
S.~Bettarini,
C.~Bozzi,
G.~Calderini,
V.~Ciulli,
R.~Dell'Orso,
R.~Fantechi,
I.~Ferrante,
A.~Giassi,
A.~Gregorio,
F.~Ligabue,
A.~Lusiani,
P.S.~Marrocchesi,
A.~Messineo,
F.~Palla,
G.~Sanguinetti,
A.~Sciab\`a,
G.~Sguazzoni,
J.~Steinberger,
R.~Tenchini,
C.~Vannini,
A.~Venturi,
P.G.~Verdini
\samepage
\begin{center}
\parbox{15.5cm}{\sl\samepage
Dipartimento di Fisica dell'Universit\`a, INFN Sezione di Pisa,
e Scuola Normale Superiore, 56010 Pisa, Italy}
\end{center}\end{sloppypar}
\vspace{2mm}
\begin{sloppypar}
\noindent
G.A.~Blair,
L.M.~Bryant,
J.T.~Chambers,
Y.~Gao,
M.G.~Green,
T.~Medcalf,
P.~Perrodo,
J.A.~Strong,
J.H.~von~Wimmersperg-Toeller
\nopagebreak
\begin{center}
\parbox{15.5cm}{\sl\samepage
Department of Physics, Royal Holloway \& Bedford New College,
University of London, Surrey TW20 OEX, United Kingdom$^{10}$}
\end{center}\end{sloppypar}
\vspace{2mm}
\begin{sloppypar}
\noindent
D.R.~Botterill,
R.W.~Clifft,
T.R.~Edgecock,
S.~Haywood,
P.~Maley,
P.R.~Norton,
J.C.~Thompson,
A.E.~Wright
\nopagebreak
\begin{center}
\parbox{15.5cm}{\sl\samepage
Particle Physics Dept., Rutherford Appleton Laboratory,
Chilton, Didcot, Oxon OX11 OQX, United Kingdom$^{10}$}
\end{center}\end{sloppypar}
\vspace{2mm}
\begin{sloppypar}
\noindent
B.~Bloch-Devaux,
P.~Colas,
B.~Fabbro,
W.~Kozanecki,
E.~Lan\c{c}on,
M.C.~Lemaire,
E.~Locci,
P.~Perez,
J.~Rander,
J.-F.~Renardy,
A.~Rosowsky,
A.~Roussarie,
J.-P.~Schuller,
J.~Schwindling,
A.~Trabelsi,
B.~Vallage
\nopagebreak
\begin{center}
\parbox{15.5cm}{\sl\samepage
CEA, DAPNIA/Service de Physique des Particules,
CE-Saclay, 91191 Gif-sur-Yvette Cedex, France$^{17}$}
\end{center}\end{sloppypar}
\vspace{2mm}
\begin{sloppypar}
\noindent
S.N.~Black,
J.H.~Dann,
H.Y.~Kim,
A.M.~Litke,
M.A. McNeil,
G.~Taylor
\nopagebreak
\begin{center}
\parbox{15.5cm}{\sl\samepage
Institute for Particle Physics, University of California at
Santa Cruz, Santa Cruz, CA 95064, USA$^{19}$}
\end{center}\end{sloppypar}
\pagebreak
\vspace{2mm}
\begin{sloppypar}
\noindent
C.N.~Booth,
R.~Boswell,
C.A.J.~Brew,
S.~Cartwright,
F.~Combley,
M.S.~Kelly,
M.~Lehto,
W.M.~Newton,
J.~Reeve,
L.F.~Thompson
\nopagebreak
\begin{center}
\parbox{15.5cm}{\sl\samepage
Department of Physics, University of Sheffield, Sheffield S3 7RH,
United Kingdom$^{10}$}
\end{center}\end{sloppypar}
\vspace{2mm}
\begin{sloppypar}
\noindent
K.~Affholderbach,
A.~B\"ohrer,
S.~Brandt,
G.~Cowan,
J.~Foss,
C.~Grupen,
G.~Lutters,
P.~Saraiva,
L.~Smolik,
F.~Stephan 
\nopagebreak
\begin{center}
\parbox{15.5cm}{\sl\samepage
Fachbereich Physik, Universit\"at Siegen, 57068 Siegen,
 Fed.\ Rep.\ of Germany$^{16}$}
\end{center}\end{sloppypar}
\vspace{2mm}
\begin{sloppypar}
\noindent
M.~Apollonio,
L.~Bosisio,
R.~Della~Marina,
G.~Giannini,
B.~Gobbo,
G.~Musolino
\nopagebreak
\begin{center}
\parbox{15.5cm}{\sl\samepage
Dipartimento di Fisica, Universit\`a di Trieste e INFN Sezione di Trieste,
34127 Trieste, Italy}
\end{center}\end{sloppypar}
\vspace{2mm}
\begin{sloppypar}
\noindent
J.~Putz,
J.~Rothberg,
S.~Wasserbaech,
R.W.~Williams
\nopagebreak
\begin{center}
\parbox{15.5cm}{\sl\samepage
Experimental Elementary Particle Physics, University of Washington, WA 98195
Seattle, U.S.A.}
\end{center}\end{sloppypar}
\vspace{2mm}
\begin{sloppypar}
\noindent
S.R.~Armstrong,
E.~Charles,
P.~Elmer,
D.P.S.~Ferguson,
S.~Gonz\'{a}lez,
T.C.~Greening,
O.J.~Hayes,
H.~Hu,
S.~Jin,
P.A.~McNamara III,
J.M.~Nachtman,
J.~Nielsen,
W.~Orejudos,
Y.B.~Pan,
Y.~Saadi,
I.J.~Scott,
J.~Walsh,
Sau~Lan~Wu,
X.~Wu,
J.M.~Yamartino,
G.~Zobernig
\nopagebreak
\begin{center}
\parbox{15.5cm}{\sl\samepage
Department of Physics, University of Wisconsin, Madison, WI 53706,
USA$^{11}$}
\end{center}\end{sloppypar}
}
\footnotetext[1]{Now at Princeton University, Princeton, NJ 08544, U.S.A.}
\footnotetext[2]{Permanent address: Kangnung National University, Kangnung,
Korea.}
\footnotetext[3]{Also at Dipartimento di Fisica, INFN Sezione di Catania,
Catania, Italy.}
\footnotetext[4]{Also Istituto di Fisica Generale, Universit\`{a} di
Torino, Torino, Italy.}
\footnotetext[5]{Also Istituto di Cosmo-Geofisica del C.N.R., Torino,
Italy.}
\footnotetext[6]{Supported by the Commission of the European Communities,
contract ERBCHBICT941234.}
\footnotetext[7]{Supported by CICYT, Spain.}
\footnotetext[8]{Supported by the National Science Foundation of China.}
\footnotetext[9]{Supported by the Danish Natural Science Research Council.}
\footnotetext[10]{Supported by the UK Particle Physics and Astronomy Research
Council.}
\footnotetext[11]{Supported by the US Department of Energy, grant
DE-FG0295-ER40896.}
\footnotetext[12]{Also at CERN, 1211 Geneva 23,Switzerland.}
\footnotetext[13]{Supported by the US Department of Energy, contract
DE-FG05-92ER40742.}
\footnotetext[14]{Supported by the US Department of Energy, contract
DE-FC05-85ER250000.}
\footnotetext[15]{Permanent address: Universitat de Barcelona, 08208 Barcelona,
Spain.}
\footnotetext[16]{Supported by the Bundesministerium f\"ur Bildung,
Wissenschaft, Forschung und Technologie, Fed. Rep. of Germany.}
\footnotetext[17]{Supported by the Direction des Sciences de la
Mati\`ere, C.E.A.}
\footnotetext[18]{Supported by Fonds zur F\"orderung der wissenschaftlichen
Forschung, Austria.}
\footnotetext[19]{Supported by the US Department of Energy,
grant DE-FG03-92ER40689.}
\footnotetext[20]{Now at School of Operations Research and Industrial
Engireering, Cornell University, Ithaca, NY 14853-3801, U.S.A.}
\footnotetext[21]{Now at Schweizerischer Bankverein, Basel, Switzerland.}
%
%
\setlength{\parskip}{\saveparskip}
\setlength{\textheight}{\savetextheight}
\setlength{\topmargin}{\savetopmargin}
\setlength{\textwidth}{\savetextwidth}
\setlength{\oddsidemargin}{\saveoddsidemargin}
\setlength{\topsep}{\savetopsep}
\normalsize
\newpage
\pagestyle{plain}
\setcounter{page}{1}

\section{Introduction}
  The Standard Model (SM) predicts the production of events 
at LEP2 with one
or more photons and missing energy through two processes: radiative
returns to the Z resonance ($ \rm e^{+} e^{-} \rightarrow \gamma  Z$)  with Z$\rightarrow \nu \bar{\nu}$, and $t$-channel W 
exchange with photon(s) radiated from the beam electrons or the W. ~\,
Such events have been studied in e$^{+}$e$^{-}$ annihilations at centre-of-mass
energies of 130 and 136 GeV~\cite{newexp} and  161 GeV~\cite{opal} as well as at previous collider experiments~\cite{oldexp}.
The process \nunug\ is well understood
theoretically, so any significant deviation from the predictions of
the Standard Model could signal new physics.  

  Events with one or more photons and missing energy could arise
in Supersymmetry where the missing energy is caused by weakly 
interacting supersymmetric particles.  For example, in the Minimal 
Supersymmetric extension of the Standard Model (MSSM), the second 
lightest neutralino can decay radiatively to the lightest neutralino~\cite{howie}.  
A large branching ratio  for this decay is expected only in a small region of parameter 
space.  This region can however be enlarged by breaking the condition on 
the unification of gaugino masses at the GUT scale.

  Alternatively there are SUSY models which postulate that the lightest supersymmetric particle (LSP) is the 
gravitino.  In these models the lightest neutralino decays to an essentially massless 
gravitino (M$_{\tilde{G}}<1 ~\, {\rm MeV}/c^{2}$) and a photon with a 100\% branching ratio.  Examples include
the so-called ``No-Scale Supergravity'' (LNZ model)~\cite{lnz} and models with gauge-mediated supersymmetry breaking (GMSB)~\cite{life,gmsbbig,dine,gmsb,fayet}.  
 In both  classes of models, one might expect the process \lepg\ at LEP2,
 seen in the detector  as two photons
 and missing energy. 
  The one-photon process \lepgx\ is expected to be produced at LEP2 for only very light gravitino masses as the cross section scales as the inverse of the gravitino mass
squared~\cite{fayet2}.  For a gravitino mass of
$10^{-5} ~\, {\rm eV}/c^{2}$  the cross section is predicted to be around 1 pb.
  In the LNZ model the gravitino mass is allowed to be this light~\cite{lnz2},
 but in GMSB the gravitino is predicted to have a mass five orders of magnitude
bigger.  Thus, this process is not expected in  GMSB.


  The data collected by ALEPH at energies of 161, 170 and 172 GeV (11.1,
1.1 \pb\ and 9.5~\pb , respectively) have
been analysed for anomalous single-photon and two-photon production
using criteria optimized for a range of SUSY particle masses.  No evidence for anomalous photon(s) and missing energy events is found,
and limits are placed on the production of supersymmetric particles
in the context of the models introduced above.   For GMSB, the neutralino composition is assumed to be pure bino throughout this letter.

 The CDF collaboration has observed an unusual event with  two high energy electrons, two high energy photons, and a large amount of missing transverse energy~\cite{cdf}.
The SM explanation for this event has a low probability, but it can be accommodated by the SUSY models mentioned above. 
In the neutralino LSP scenario the CDF event could be explained by the Drell-Yan process 
\cdfx\
where the two \xone 's escape detection resulting in missing transverse 
energy.  If this is the explanation for the CDF event, the best possibility for discovery at LEP2 is \lepx .
In principle \lx\ could be considered, however the predicted cross section
is uninterestingly small.
In gravitino LSP models, the CDF event could be explained by
\cdfg .
  In this scenario the best channel for discovery at LEP2
 is \lepg .
 The limits derived from the ALEPH data are compared to the
regions favored by  the CDF event within these models.

  The outline of this letter is as follows: after a brief description of the detector in Section 2, the Monte Carlo samples are presented in Section 3, the
 one photon and two photon plus missing energy searches are detailed in Sections 4 and 5 and conclusions are stated in Section 6.

\section{The ALEPH detector and photon identification}

The ALEPH detector   and its performance are described in detail elsewhere \cite{ALEPH_PAPER,PERFORMANCE}. 
The analysis presented here depends largely on the performance of the electromagnetic 
calorimeter (ECAL). 
The luminosity calorimeters (LCAL and SICAL), together with the hadron calorimeter (HCAL), 
 are used mainly to veto events in which photons are accompanied by other energetic 
particles. 
 The HCAL is instrumented with streamer tubes, which are useful in identifying
 muons.
The SICAL  provides coverage between 34 and 63 mrad from the beam axis while the LCAL provides coverage  between 45 and 160 mrad.  The LCAL consists of two halves which fit together around the beam axis; the area where the two halves join is a region of reduced sensitivity ('the LCAL crack').
 This vertical crack accounts for only 0.05\% of the total solid angle coverage of the ALEPH detector.
The tracking system, composed of a silicon vertex detector, wire drift chamber (ITC), 
and time projection chamber (TPC), is used to provide efficient
($>99.9\%$) tracking of isolated charged particles in the angular range
$|\cos\theta| < 0.96$.

The ECAL is a lead/wire-plane sampling calorimeter consisting of 36 modules, twelve in the barrel and twelve 
in each endcap, which provide coverage in the angular range ${|\cos\theta|\,<\,0.98}$.
Inter-module cracks reduce this solid angle coverage by 2\% in the barrel and 6\%
in the endcaps.   However, the ECAL  and HCAL cracks are not aligned so there is complete coverage in ALEPH at large polar angles.
 At normal incidence the ECAL is situated at 185 cm from the interaction point.
The total thickness of the ECAL is 22 radiation lengths at normal incidence.
Anode wire signals, sampled every $512\,\mathrm{ns}$ during their rise time, provide 
a measurement by the ECAL of the interaction time $t_{0}$ of the particles relative to 
the beam crossing with a resolution better than $15\,\mathrm{ns}$ (for showers with energy greater than 1 \gev ). 
Cathode pads associated with each layer of the wire chambers are connected to form 
projective `towers', each subtending approximately $0.9^{\circ}\times0.9^{\circ}$,
which are oriented towards the interaction point. 
Each tower is read out in three segments in depth of four,
nine and nine radiation lengths. 
The high granularity of the calorimeter provides excellent
identification of photons and electrons. 
The energy calibration of the ECAL is obtained
from Bhabha and gamma-gamma events. 
The energy resolution is measured to be 
$\Delta{E/E}~=~0.18/\sqrt{E}~+~0.009$ ($E$ in GeV) \cite{PERFORMANCE}.

Photon candidates are identified using an
algorithm \cite{PERFORMANCE} which
performs a topological search for localised energy depositions within 
groups of neighboring ECAL towers.
These energy depositions are required to have transverse and 
longitudinal profiles consistent with that of an electromagnetic shower. 
 Photons  far from ECAL cracks have their energy measured solely from the localised energy deposition.
 In order to optimise the energy reconstruction, photons that are not well-contained in the ECAL (near or in a crack) have their energy measured from the sum of the localised energy depositions and all energy deposits in the  HCAL within a cone of \cosa . 
Photon candidates may also be identified in the tracking system  
if they convert producing an electron-positron pair~\cite{PERFORMANCE}.

The trigger most relevant for photonic events is the neutral energy trigger.
For the 1996 run, the total wire energy measured in an ECAL module must be greater than 1 \gev\ in the barrel and 2.3 \gev\  in the endcaps in order for this trigger to fire.  The neutral energy trigger is fully efficient for the analyses to be described.  

\section{The Monte Carlo samples}
 The efficiency for the \eenunu\ cross section  measurement and the background for the anomalous photon plus missing energy searches are estimated using the KORALZ  Monte Carlo program~\cite{koralz}.
This Monte Carlo is  checked by comparing to NUNUGG~\cite{nngg} at \ecm\ below the W threshold and to CompHEP~\cite{comphep}  at higher energies.
The Monte Carlos agree within errors to 1\%  for the emission of one photon.   The two photon plus missing energy signature is checked for loose acceptance cuts (E$_{\gamma} > 5 \, \rm GeV$, $ \left|\cos\theta\right| < 0.95$) and in a more restrictive region 
($\rm Missing \, Mass > 100 \, GeV/ {\mathit c}^{2}$).  The discrepancy between KORALZ and CompHEP is at most 10\%~\cite{gam2} .

Background to the photon(s)  plus  missing energy signature  can come from \eegg\ and \bha\
where initial or final state particles radiate a photon and the final state
particles escape along the beam direction undetected.
  This background is studied using the GGGB03~\cite{gggb} and BHWIDE~\cite{bhwide}  Monte Carlo programs, respectively.
 The signal generator SUSYGEN~\cite{susygen} is used to design the selection criteria and evaluate the efficiency for the searches for new physics.

\section{ One photon and missing energy }

 The Standard Model predicts a large cross section for the process \eenunu , which  constitutes an  irreducible background to searches for new physics.   One can, however, still search for new physics by  observing an excess of events over the  Standard Model prediction.
 This requires a precise understanding of the background level from \eenunu\ and a reduction of cosmic ray and detector noise events (non-subtractable backgrounds) to a negligible level.

\subsection{Event selection}

Initially, events  are selected with no charged tracks (not coming from a conversion) and exactly one photon inside the acceptance cuts of
 ${\mathrm | \cos{\theta} | < 0.95}$ and $p_{\perp} >$ \ptac\ (where $p_{\perp}$ is defined as the measured transverse momentum relative to the beam axis). 
The remaining selection criteria are tailored to eliminate as much as possible the non-subtractable backgrounds.
Cosmic ray events that traverse the detector are eliminated  by the charged track requirement or if there are hits in the outer part of the HCAL.
 A small fraction of cosmic ray events and  detector noise events in the ECAL 
remain after these cuts.  The ECAL information can be exploited to remove these types of events.
 The barycentre of the photon shower is found in each of the three ECAL stacks.  Taking two points at a time, three possible photon trajectories are calculated and used to estimate the distance of closest approach of the photon to the interaction point.  The smallest of the three distances
(`impact parameter of the photon')
is required to be less than 25 cm.  
The compactness of the shower in the ECAL is calculated by taking an energy-weighted average of the angle subtended at the interaction point
between the cluster barycentre and the barycentre of each of the ECAL storeys 
contributing to the cluster.  The compactness is required to be less than 
0.85$^{\circ}$.   
 Both the impact parameter and compactness distributions  are peaked at
low values for photons coming from the interaction point and at high values for the remaining background.
The cuts are chosen so that there is negligible efficiency loss for real photons.
 Finally, the interaction time is required to be within 
40 ns of the beam crossing.

 A single photon and missing energy might also be caused by initial or final state radiation when the final state particles escape along the beam direction.
 Most of the events that satisfy the requirement that the $p_{\perp}$ of the photon must be greater than \ptac\
  will have a  particle
 within the angular acceptance of the detector.
These
events are eliminated by  the no charged track requirement and the
 following cut. 
The event is rejected if the total
 energy  in the detector excluding the photon is more 
than 1 \gev\ 
 or if there is any energy reconstructed within 14$^{\circ}$
of the beam axis.   This requirement reduces the efficiency for
\eenunu\
 by 11\% due to additional brehmsstrahlung photons at low angles (7\% loss) and uncorrelated noise in the detector (4\% loss).
To compensate for vertical cracks in the LCAL,
 the $p_{\perp}$ requirement is tightened to 0.145$\sqrt{s}$ if the missing  momentum vector points to within 17$^{\circ}$ of 90$^{\circ}$ or 270$^{\circ}$ in azimuthal angle ($\phi$). 

 Residual cosmic ray  and detector noise backgrounds are measured by selecting events slightly out of time with respect to the beam crossing but which pass all other cuts.
No such events are found in a displaced time window of 740 ns width.  
Events with a radiated photon in the acceptance  and the final state particles escaping undetected along the beam axis are studied using Monte Carlo.
The equivalent of 60 data sets  of these SM background processes was generated and passed through the full detector simulation.   No events from these samples survive the selection.

\begin{table}
\begin{center}
\begin{tabular}{|l|c|c|}
\hline
 & \multicolumn{2}{c|}{ Efficiency(\%)} \\ \cline{2-3}
Selection        & 161 GeV & 172 GeV \\ \hline
N$_{\gamma} \geq$ 1 and N$_{\rm ch}$ = 0           & 94 & 94 \\
N$_{\gamma}$ = 1                    & 89 & 89 \\
$p_{\perp}$ $> 0.145 \sqrt{s} \, \rm if \, \phi_{pmiss} = 90 \pm 17 \, or \, 270 \pm 17$ & 82 & 82 \\
Additional energy  $<$ 1 \gev\                   & 72 & 74 \\
No energy within 14$^{\circ}$ of the beam axis         & 71 & 73 \\  
All other cuts                        & 70 & 72 \\ \hline
\end{tabular}
\end{center}
\caption {{\sl The cumulative efficiency for the \eenunu\ process inside the acceptance cuts. 
}}
\label{effew}
\end{table}
\subsection{Measurement of the \eenunu\ cross section}
The efficiency for the process \eenunu\ is detailed in Table~\ref{effew}. 
The
efficiency loss due to uncorrelated noise or beam-related background in the detector is estimated using events triggered at random beam crossings and is included  in the efficiency estimate.  
 Applying the selection criteria to the data, 41 one-photon events are
found at $\sqrt{s}$ = ~161 \gev\ while 45 are expected from Monte Carlo.  At $\sqrt{s}$ = 170/172 GeV, 36 one-photon events are found while 37 are expected.

\begin{figure}[t]
\epsfxsize=15cm
\epsfysize=15cm
\begin{center}\mbox{\epsfbox{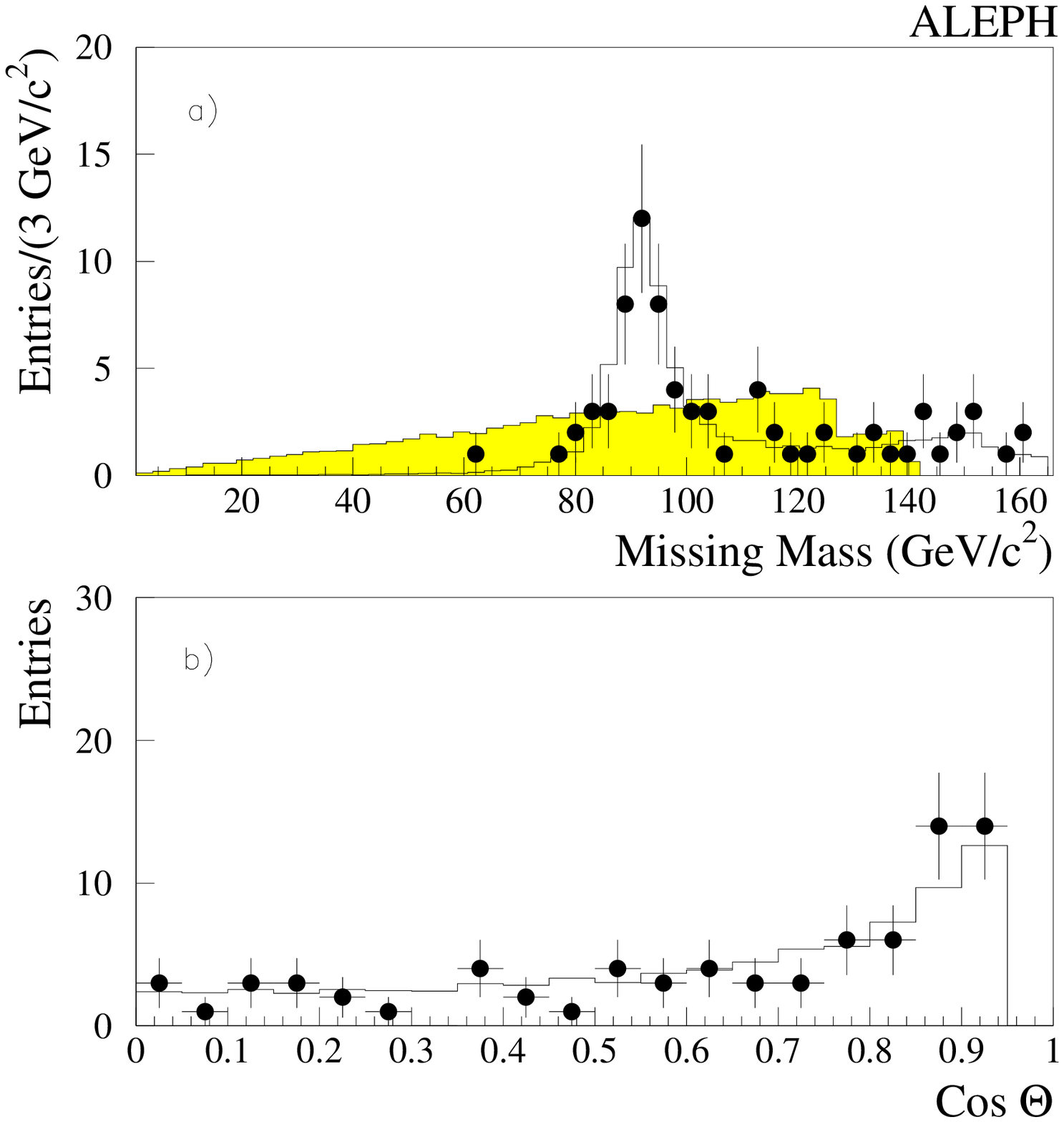}}\end{center}
\caption{{\sl
a) The invariant mass distribution of the system recoiling against the photon 
candidate is shown for both the data (points with error bars) and Monte Carlo (histogram).  The signal \lepgx\  for a \xone\  mass of 100 \gevsq\ is overlaid (shaded histogram) with arbitrary normalization.
b)   The \cs\ distribution is shown for both the data (points with error bars) and Monte Carlo (histogram).  The signal has a flat distribution in \cs .
}}
\label{gmass}
\end{figure}

Inside the acceptance $| \cos{\theta} | < 0.95$ and $p_{\perp} >$ \ptac\ the  cross section measurements are

\[\mathrm \sigma(e^{+}e^{-}\rightarrow   \nu \bar{\nu} \gamma (\gamma)) = 5.3 \pm 0.8 \pm 0.2  \, pb \hspace{10mm} \sqrt{s}=161 \, \gev\ \]
\[\mathrm \sigma(e^{+}e^{-}\rightarrow   \nu \bar{\nu} \gamma (\gamma)) = 4.7 \pm 0.8 \pm 0.2  \, pb \hspace{10mm} \sqrt{s}=172 \, \gev .  \]
%
These results are consistent with the Standard Model predictions
of 5.81 $\pm$ 0.03 pb at 161 \gev\ and 4.85 $\pm$ 0.04 pb at 172 GeV obtained using
the KORALZ Monte Carlo.
The missing mass  and polar angle distributions are shown in Figure~\ref{gmass}.

 The estimates of the systematic uncertainties in the above cross sections include contributions from the sources listed in Table~\ref{systew}.
The simulation of the energetic photon shower is checked with a sample of Bhabha events selected  requiring two collinear beam-momentum tracks and using muon
chamber information to veto $\mu^{+}\mu^{-}$ events.  The tracking information was masked from these events and the photon reconstruction redone.  The efficiency to reconstruct a photon in these events is found to be consistent  within
the available statistics at the 3\% level. The uncertainty in the number of simulated pair conversions is estimated to give a 0.3\% change in the overall efficiency. 
To account for the uncertainty in the energy calibration the energy is shifted by 
2\% and the efficiency is recalculated. The difference in the efficiency is found to be 0.2\%.  
  The total systematic uncertainty is obtained by adding in quadrature the individual contributions.
\begin{table}
\begin{center}
\begin{tabular}{|l| |r|}
\hline
Source  & Error(\%) \\ \hline
Photon selection                               & 3\\
Converted photon selection                     & 0.3 \\
Energy calibration                 & 0.2 \\
Background                         & $<$1 \\ \hline
Integrated luminosity                          & 0.7 \\ \hline
Monte Carlo theoretical                        & 1 \\
Monte Carlo statistical                        & 0.4 \\ \hline \hline
Total (in quadrature)                          & 4 \\ \hline
\end{tabular}
\end{center}
\caption {{\sl Systematic uncertainties for the one-photon channel. 
}}
\label{systew}
\end{table}
\subsection{Search for a light gravitino in the one-photon channel}

In order  to  search for the signal \lepgx , a binned maximum likelihood fit is  performed on the  observed missing mass spectrum under the hypothesis that there is a mixture of signal and background  in the data. 
  Events from the \eenunu\  and \lepgx\ processes have very different  missing mass distributions, as shown in Figure~\ref{gmass}. 
The likelihood that the missing mass distribution of the data agrees with the composite missing mass distributions of the Monte Carlo background \eenunu\ and signal \lepgx\ processes is calculated. 
Following the method of Ref.~\cite{cl}, the upper limit on the total number of signal events $S$ is calculated by integrating the likelihood  as a function of $S$. The number of expected signal events is increased until the integration from $S=0$ to $\mu_{S}$ is 95\% of the total area (the integration from $S=0$ to $\infty$). The upper limit on the total number of signal events at the 95\%  confidence level is then given by $\mu_{S}$.  This procedure is repeated at each neutralino mass ranging from 40 \gevsq\ to 171 \gevsq\ in steps of 1~\gevsq .

 A toy Monte Carlo with the kinematic cuts applied is used to describe the signal shape of the missing mass distribution for each neutralino mass. 
The MC is used to estimate the efficiency loss due to initial state radiation and photon reconstruction.  The efficiency loss due to noise in the detector is
also included.

\begin{figure}[t]
\epsfxsize=10cm
\epsfysize=10cm
\begin{center}\mbox{\epsfbox{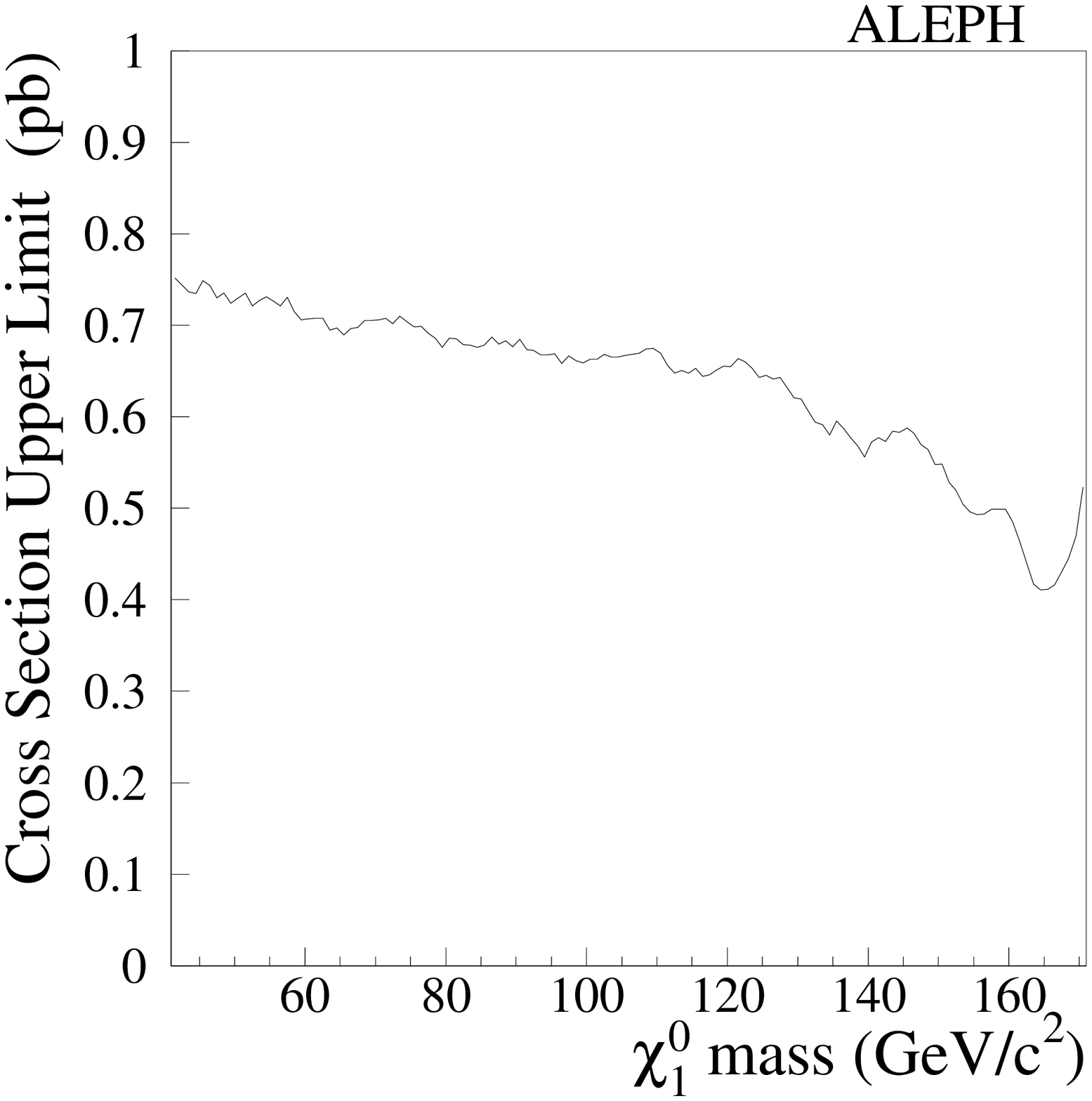}}\end{center}
\caption{{\sl
The 95\% C.L. upper limit on the production cross-section for
\lepgx .
The limit is valid for \ecm\ = 172 \gev\ assuming $\beta^{8}$  threshold dependence
 and isotropic decays. 
}}
\label{explim1}
\end{figure}

 The  upper limit on the cross section at 95\% confidence level is shown in Figure~\ref{explim1}. A negligible neutralino lifetime is assumed. 
  The luminosity of the two data samples is combined assuming $\beta^{8}$  threshold dependence of the cross section.
 The systematic uncertainty is taken into account following Ref.~\cite{syst},
 which changes the upper limit on the number of signal events by less than 1\%.
 In the LNZ theory~\cite{lnz2},
 for a gravitino mass of 10$^{-5}$ eV/$c^{2}$, the mass limit  for the neutralino is
 100 \gevsq . 
However, the cross section for this
process scales as the inverse of the gravitino mass squared, so
the limit on the neutralino mass is very sensitive to the assumed
gravitino mass.

\section{Two photons and missing energy}
  As described in the introduction, there are two SUSY scenarios
which can give acoplanar photons: the gravitino LSP and
neutralino LSP scenarios.  The signals differ in that the
invisible particle is essentially massless in the first scenario
and can have substantial mass in the second one.  This leads to two
slightly different search criteria, as described in the subsections
below.

  The cross section for the SM background process \nunugg\
is reduced by order $\alpha$ from the single-photon cross 
section, so a cut-based analysis is sufficient to search for new physics.
 The preselection begins by requiring no charged tracks that do not come from
a conversion.  Due to detector acceptance only photons within ${\mathrm | \cos{\theta} | < 0.95}$ are counted.
 Since at least two photons are required, background from cosmic rays and 
detector noise is less severe, so the impact parameter 
requirement is not imposed.
 Events with more than two photons are required to have at least $0.4 \sqrt{s}$ of missing energy.
Missing transverse energy is required by imposing an acoplanarity cut  at 177$^{\circ}$ and requiring that the additional total energy be less than 1 \gev . 
  When there are three or more photons in the event, the two most energetic photons are used to determine the acoplanarity.
 The \eegg\ background is effectively eliminated after these selection criteria. The total $p_{\perp}$ is required to be greater than 3.75\% of the missing energy, reducing background from radiating events   with final state particles escaping down the beam-axis to a negligible level.  
The selection criteria are shown in Table~\ref{acoplanar}.
\begin{table}
\begin{center}
\begin{tabular}{|l|c|c|c|}
\hline
      &Cumulative  & \nunuggg\ bkg. &  \ggg\ bkg. \\
Two-photon selection criteria  & signal eff.(\%) & $\sigma$ (pb) & $\sigma$ (pb) \\ \hline
N$_{\gamma}$=2 OR (N$_{\gamma} \geq 3 \, \rm and \, E_{\rm missing} > 0.4 \sqrt{s}$)   & 83 & 0.36  & 11.9   \\ 
Acoplanarity $< 177^{\circ}$ & 81 & 0.35   & 0.3   \\  
Additional energy $<$ 1 GeV & 73 & 0.32   & 0.008   \\
Total $p_{\perp} > 0.0375 \ast \rm E_{missing}$ & 73 & 0.30   & 0.002   \\ \hline
\rule{0pt}{14pt} 
\underline{$\tilde{G}$ LSP analysis} & & & \\
E$_{2}$ $\geq$ 18 \gev\ & 69 & 0.043   & 0.002   \\ \hline 
\rule{0pt}{14pt} 
\underline{\xone\ LSP analysis} & & & \\
M$_{\rm missing} \leq$ 82 GeV/c$^{2}$ OR M$_{\rm missing}  \geq$ 100 GeV/c$^{2}$ & 71 & 0.16   & 0.002   \\
OR E$_{2} \geq$ 10 \gev\  &   &  &  \\
Two photons inside \cost2\ & 52 & 0.063   & - \\ \hline 

\end{tabular}
\end{center}
\caption {{\sl Two-photon selection criteria, and the additional cuts required by the two analyses described in the text.  Signal efficiency for the gravitino LSP analysis is given for a 65 \gevsq\ \xone\ at $\sqrt{s} = 161 \, \rm GeV$.  For the \xone\ LSP analysis the efficiency numbers are given for a 45 \gevsq\ \xtwo\ and a 20 \gevsq\ \xone .  Background numbers are given for 
$\sqrt{s}$ = 161 \gev\ but are similar for 172~GeV. 
}}
\label{acoplanar}
\end{table}
  After this initial selection, two events are selected at 161 GeV
while 2.7 are expected from \eenunu .  At 172~GeV, one event is selected while 2.3
are expected.

\subsection{Acoplanar photon search: gravitino LSP scenario}

\begin{figure}[t]
\epsfxsize=10cm
\epsfysize=10cm
\begin{center}\mbox{\epsfbox{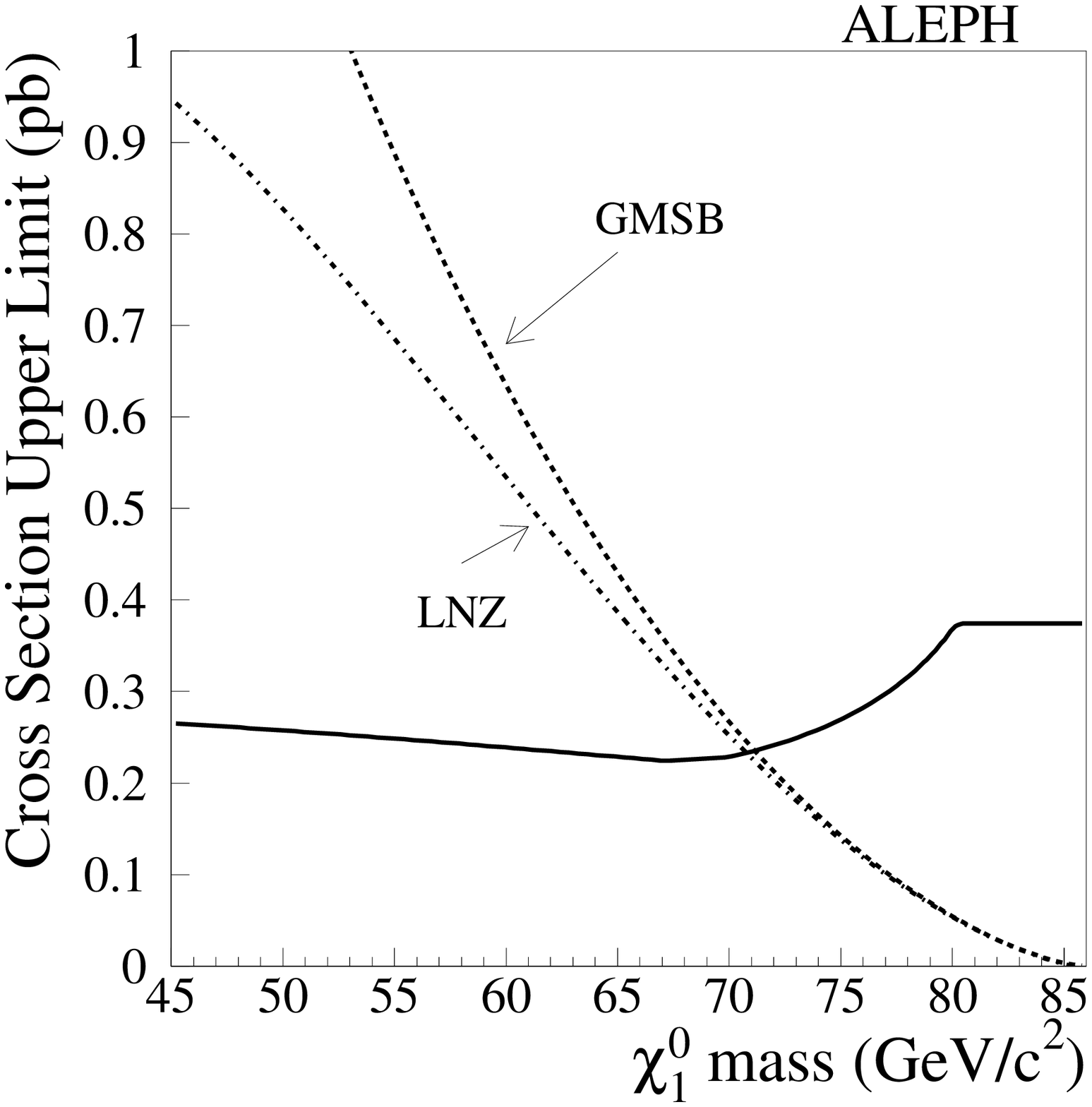}}\end{center}
\caption{{\sl
The 95\% C.L. upper limit on the production cross section for 
\lepg\ when
\xone\ has a lifetime less than 3 ns.  The limit is valid for \ecm\ = 172 \gev . The data from 161~GeV are included by scaling the luminosity by the ratio of the cross section at that energy to the cross section at 172 \gev .  Two different theories are compared to the experimental limit. The right selectron mass is taken to be 1.5 that of the neutralino mass for the GMSB Theory.
}}
\label{explim}
\end{figure}

   An additional cut is placed on the energy of the less energetic
photon (E$_{2}$) to reduce substantially the remaining SM background.  The
energy distribution of the second  most energetic photon is peaked near zero 
for the background,  whereas for the signal both photons have a flat distribution in an interval
depending on the neutralino mass and \ecm .  
 The theoretical cross section for a pure bino neutralino is used to estimate the expected limit on the neutralino mass.   The cut is placed according to the
\={N}$_{95}$ optimisation procedure~\cite{janot}
 at E$_{2} < 18 \, \rm GeV$, reducing the background to 43 fb,
while the efficiency remains high at 69\% for a neutralino of 65 \gevsq\ mass produced at 161~GeV. 
 After this selection criteria no events are found in the data  while 0.92 events are expected from background processes.  
Figure~\ref{explim}  shows the upper limit on the cross section   
compared to two theoretical predictions.  
 The integrated luminosity taken at \ecm\ = 161 \gev\  is scaled by the ratio
of cross sections to those at 172 \gev . The neutralino is taken to be pure bino and the right-selectron mass is set to
1.5 the neutralino mass.
 The upper limit on the cross section is not strongly dependent on the above
choices.  Scaling the luminosity at 161 \gev\ by the threshold dependence
$\beta^{3}/s$ changes the  cross section limit by less then 5\%.
The mass limit obtained  for both the GMSB and LNZ models is 
\vspace{.5cm}
\begin{center}
\large
$ \mathrm M_{\chi_{1}^{0}} \geq 71 \,  GeV/c^{2}  $  
\end{center}
\vspace{.5cm}
\normalsize
 at 95\% C.L. for a neutralino with $\tau_{\chi_{1}^{0}} \leq $ 3 ns.
 The analysis is efficient as long as the \xone\  decays inside the ECAL. 
%
The systematic uncertainty for this analysis is less than 6\%, dominated by photon reconstruction efficiency.  The effect of this uncertainty on the  cross section upper limit is less then 1\%, taken into account by means of the method of Ref.~\cite{syst}.
 The effect on the mass limit is negligible.

\begin{figure}[t]
\epsfxsize=10cm
\epsfysize=10cm
\begin{center}\mbox{\epsfbox{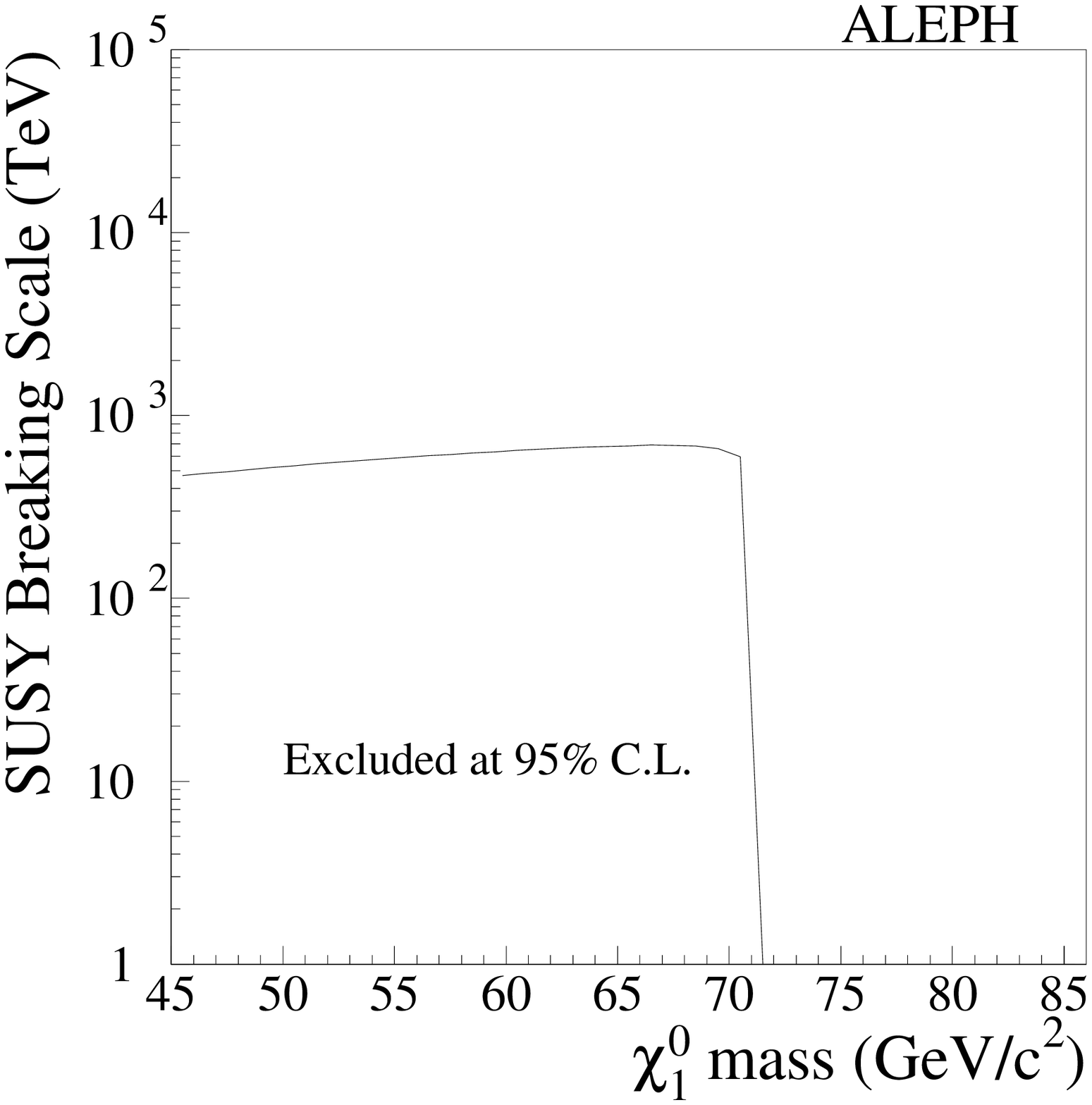}}\end{center}
\caption{{\sl
The excluded region in the neutralino mass, $\sqrt{F}$ plane, where the selectron mass is set to 1.5 times the neutralino mass and the neutralino composition is pure bino.
}}
\label{glimlife}
\end{figure}

In the GMSB model the neutralino can have a non-negligible lifetime which
depends directly on the SUSY breaking scale $\sqrt{F}$. 
The lifetime of the neutralino is given by~\cite{life}
\vspace{.5cm}
\begin{center}
\large
$\mathrm c \tau \simeq 130 \left (\frac{100 \, GeV/c^{2} }{M_{\chi_{1}^{0}}} \right)^5 
\left(\frac{\sqrt{F}}{100 \, TeV}\right)^4 \mu m $.
\end{center}
\vspace{.5cm}
\normalsize
For a neutralino of mass 71 \gevsq\ and lifetime  3 ns, the 
SUSY breaking scale is 600 TeV. 
Figure~\ref{glimlife} 
shows the 95\% C.L. exclusion limit  in the $\sqrt{F}$, M$_{\chi_{1}^{0}}$ plane.

At LEP2 the production of bino neutralinos would proceed via $t$-channel selectron exchange.  Right-selectron exchange dominates over left-selectron exchange.  Thus, the cross section for \lep1\
depends strongly on the right-selectron mass.  The theoretical cross section for \lep1\ is calculated at each M$_{\tilde{e}_{R}}$, M$_{\chi_{1}^{0}}$ mass point for right-selectron masses ranging from 70 \gevsq\ to $200 ~\, {\rm GeV}/c^{2}$  and neutralino masses ranging from 30 \gevsq\ to 86 \gevsq\  and compared to the experimental limit to obtain the exclusion region.
The neutralino mass limits were also checked for various left-selectron masses.
The result is found to be robust at the $\pm$1 \gevsq\ level for left-selectron masses ranging from 
M$_{\tilde{e}_{L}}$ = M$_{\tilde{e}_{R}}$ to
M$_{\tilde{e}_{L}} \gg$M$_{\tilde{e}_{R}}$. 

\begin{figure}
\epsfxsize=12cm
\epsfysize=12cm
\begin{center}\mbox{\epsfbox{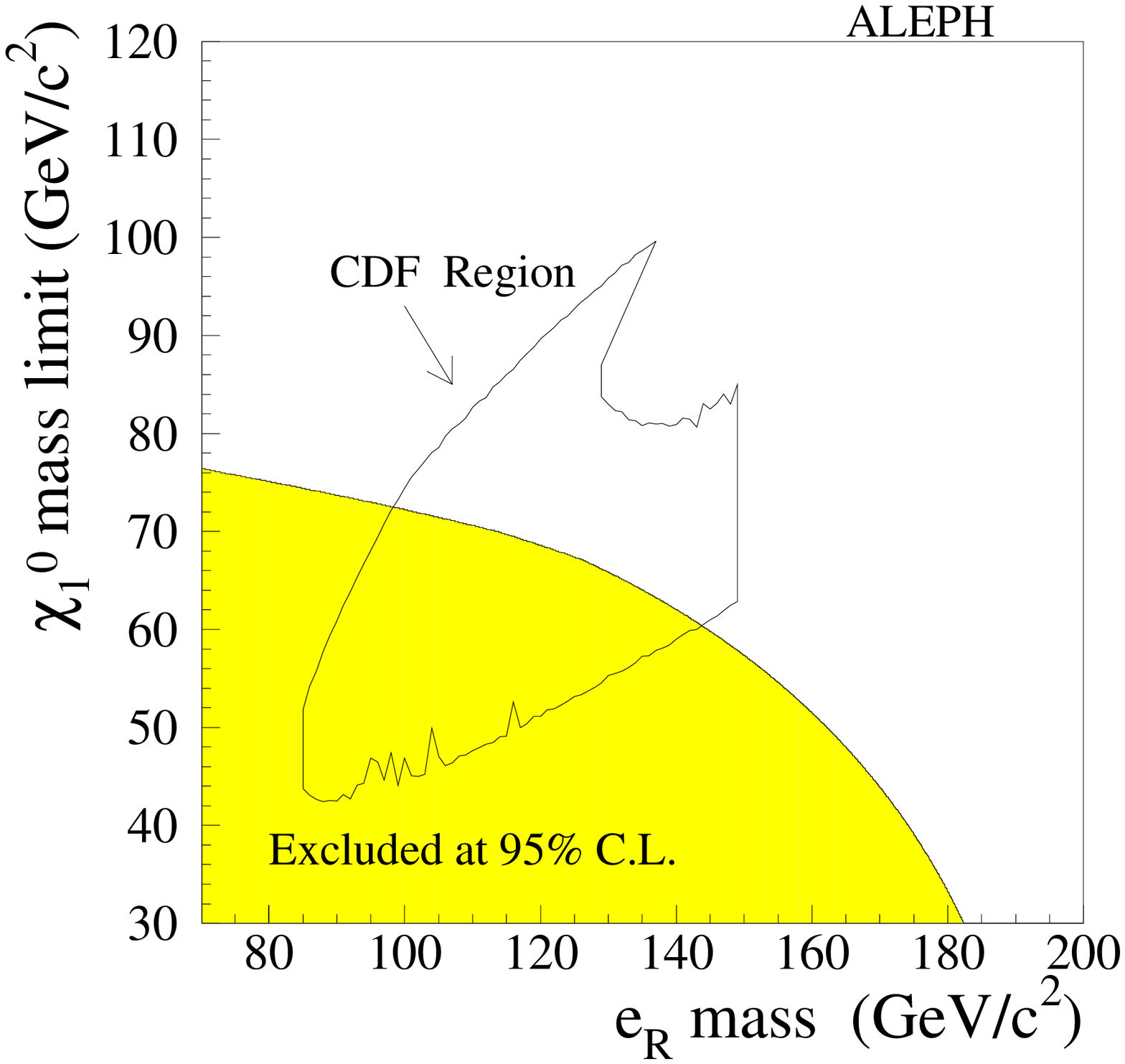}}\end{center}
\caption{{\sl The  excluded region  in the neutralino, selectron mass plane at 95\% C.L. for a pure bino neutralino (shaded area).
  Overlayed is the CDF region determined from the properties of the CDF event assuming  the reaction 
\cdfg\ (taken from the Ref.~\cite{lnz}).
}}
\label{cdfexcl}
\end{figure}

The experimentally excluded region in the neutralino, selectron mass plane is shown in Figure~\ref{cdfexcl}.  Overlayed is the 'CDF region', the area in the neutralino, selectron mass plane where the properties of the CDF event are compatible with the process 
\cdfg .
Half of the CDF region is excluded at 95\% C.L. by this analysis.

\subsection{Acoplanar photon search: neutralino LSP scenario}
     For the neutralino LSP scenario, a simple energy cut is not optimal since the \xone\ is massive and the photons from the \xxg\ decay can have low energy.  Here the fact that the \nunuggg\ background peaks at small polar angles and has a missing mass
near the Z mass is utilised. 
Events that have missing mass between 82 \gevsq\ and 100 \gevsq , and the energy of the second most energetic photon less than 10 GeV are rejected. The \cs\ cut is optimised by  using the \={N}$_{95}$ procedure, leading to a requirement of \cost2 .
 The efficiency for various \xtwo\ and \xone\  masses is shown in Table~\ref{effx2}. 
%
\begin{table}
\begin{center}
\begin{tabular}{|l|c|c|c|c|}
\hline
 & \multicolumn{4}{c|}{ $\rm{M}_{\chi_{2}^{0}-\chi_{1}^{0}} \,({\rm GeV}/c^{2})$} \\ \cline{2-5}
 $\rm{M}_{\chi_{2}^{0}} \, ({\rm GeV}/c^{2})$      & 5  & 10  & 20  & 40  \\ \hline
5 &  33  & & & \\
10 &  41  &  45  & & \\
40 &  40  &  51  & 51 & 52 \\
80 &  34  &  47  &  55  &  57 \\ \hline
\end{tabular}
\end{center}
\caption {{\sl The efficiency(\%) for the \lepx\ process at $\sqrt{s} = 161 \, \rm GeV$ . The efficiencies at 172 \gev\ are equal to (within errors)
those at 161 \gev .  
}}
\label{effx2}
\end{table}

\begin{figure}[t]
\epsfxsize=12cm
\epsfysize=12cm
\begin{center}\mbox{\epsfbox{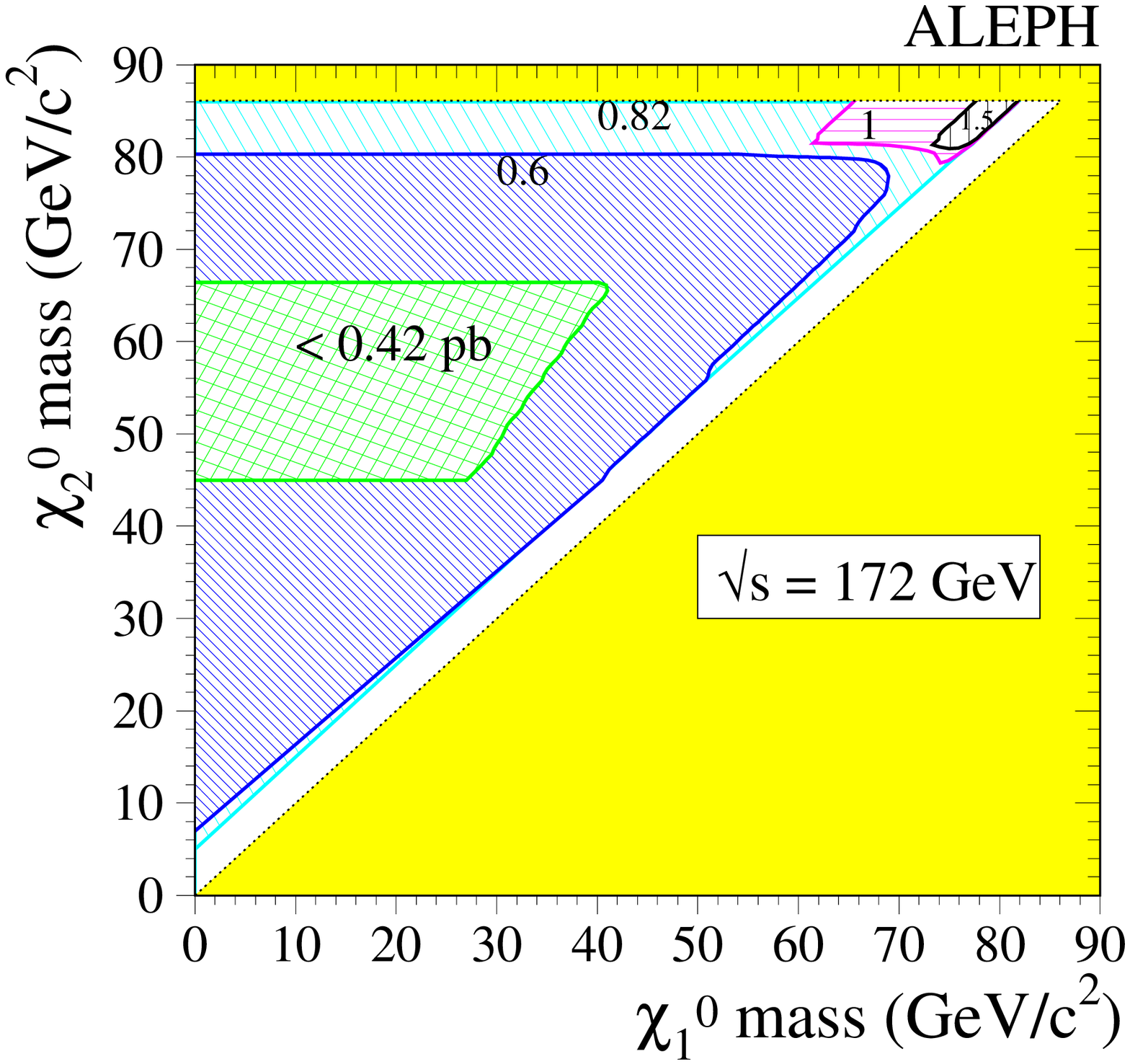}}\end{center}
\caption{{\sl
The 95\% C.L. upper limit on the production cross section for \lepx\ 
multiplied by the BR(\xxg ) squared.
 The limit is valid for \ecm\ = 172 \gev\ assuming $\beta$ threshold behavior and isotropic decays.
}}
\label{x2}
\end{figure}

One event is found in the data while 1.3 events are expected from background.  The  upper limit on the cross section in the \xone , \xtwo\ mass plane 
 are shown in Figure~\ref{x2}, assuming a branching ratio for \xxg\ of 100\%.
The systematic uncertainties for this analysis are the same as for the gravitino LSP scenario and
the effect on the upper limit  is again less than 1\%.

\begin{figure}
\epsfxsize=15cm
\epsfysize=15cm
\begin{center}\mbox{\epsfig{file=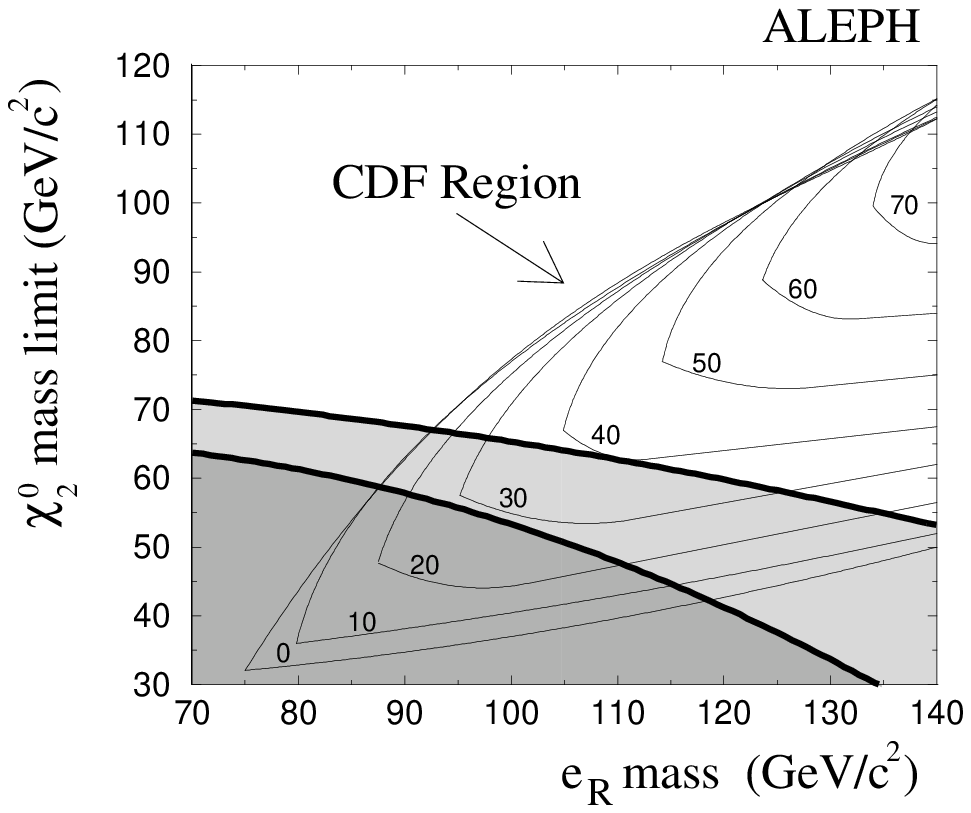,bbllx=230pt,bblly=170pt,
                    bburx=525,bbury=425,clip=}}\end{center} 
\caption{{\sl  The excluded region in the neutralino, selectron mass plane
 at 95\% C.L. 
 The \xtwo\ is pure photino  and the \xone\ is pure higgsino  which implies
 BR(\xxg ) = 1.
The shaded area is for  ${\rm M}_{\tilde{e}_{L}} = {\rm M}_{\tilde{e}_{R}}$.
 The darker shaded region refers to M$_{\tilde{e}_{L}} \gg $M$_{\tilde{e}_{R}}$. 
The mass limit is independent of the \xone\ mass as long as $ \Delta M \geq$ 25 \gevsq .
Overlayed is the CDF region  labeled by the mass of \xone\ in \gevsq . This is the area determined from the properties of the CDF event assuming 
 the reaction 
\cdfx\ (taken from Ref.~\cite{kane}).
 }}
\label{exclkane}
\end{figure}

The \xone\ LSP interpretation of the CDF event (along with the non-observation of other SUSY signatures at Fermilab) suggests a high branching ratio for
\xxg . 
 A 100\% branching ratio is achieved when the \xtwo\  is pure photino and the
\xone\ is pure higgsino.
 Assuming this scenario,  the lower mass limit of \xtwo\ as a function of the selectron mass is calculated and
compared to the region compatible with the CDF event.
In Figure~\ref{exclkane} two scenarios M$_{\tilde{e}_{L}}$ = M$_{\tilde{e}_{R}}$ 
and M$_{\tilde{e}_{L}} \gg $ M$_{\tilde{e}_{R}}$ are shown.
  With the assumption that the \xtwo\ is pure photino and the \xone\ is pure
higgsino, 
 these results exclude a significant portion of the region compatible with the  kinematics of the CDF event given by the neutralino LSP interpretation.
%
%
\section{Conclusion}

Data recorded with the ALEPH detector at  LEP centre-of-mass energies of 161 \gev\ and 170/172 \gev\ show no signs of new physics in the photon(s)  plus missing energy channels.  The cross sections and distributions for \eenunu\ 
are measured and found to be in agreement with Standard Model expectations.
The experimental 95\% C.L. upper limits on the  cross sections are derived for the following supersymmetric processes 
\lepgx , \lepg\ and \lepx .
These  cross section limits are actually more general and can be applied to the reactions:
 $\rm e^{+} e^{-} \rightarrow XY \rightarrow YY \gamma$ where Y is massless and $\rm e^{+} e^{-} \rightarrow XX \rightarrow YY \gamma \gamma$ where 
Y is massless 
or has mass.
 The 95\% C.L. limit on the \xone\ mass is found to be
71 \gevsq\ ($\tau_{\chi_{1}^{0}} \leq $ 3 ns)
 for gravitino LSP SUSY scenarios. 
 The excluded region of the SUSY Breaking Scale as a function of neutralino mass is derived.  The lower limit on the \xone\ (\xtwo ) mass as a function of selectron mass is determined and compared to the region compatible with the CDF event for the gravitino (neutralino) LSP scenario.
%
\section*{Acknowledgements}
 Fruitful discussions  with Gian Giudice are  gratefully acknowledged.
We wish to congratulate our colleagues in the CERN accelerator divisions
for the successful
 startup of the LEP2  running.
We are grateful to the engineers and technicians in all our institutions
for their contribution towards the excellent performance of ALEPH.
Those of us from non-member countries thank CERN for its ~hospitality.

\end{document}